\documentclass[11pt, a4paper]{article}
\usepackage{amsmath}
\usepackage{amssymb}
\usepackage[dvips]{graphics}
\usepackage[dvips]{graphicx}
\usepackage{soul}

\usepackage{amsmath}
\usepackage{helvet}
\usepackage{courier}
\usepackage{amssymb}
\usepackage{amsthm}
\usepackage[T1]{fontenc}
\usepackage[latin9]{inputenc}
\usepackage{makeidx}         
\usepackage{graphicx}        
\usepackage[bottom]{footmisc}

\usepackage{hyperref}
\hypersetup{colorlinks=true,
  linkcolor=blue,
  anchorcolor=blue,
  citecolor=red,
  urlcolor=brown,
  pdfauthor={Adriano Barra and Antonio Moro}}

\providecommand{\be}{\begin{equation}}
  \providecommand{\ee}{\end{equation}}
\providecommand{\bea}{\begin{eqnarray}}
  \providecommand{\eea}{\end{eqnarray}}
\providecommand{\beas}{\begin{eqnarray*}}
  \providecommand{\eeas}{\end{eqnarray*}}

\providecommand{\beni}{\begin{equation*}}
  \providecommand{\eeni}{\end{equation*}}

\providecommand{\bw}{\begin{widetext}}
  \providecommand{\ew}{\end{widetext}}

\newcommand{\benumerate}{\begin{enumerate}}
\newcommand{\eenumerate}{\end{enumerate}}

\newcommand{\der}[2]{\frac{\partial #1}{\partial #2}}

\newcommand{\av}[1]{\langle #1 \rangle}

\date{}

\begin{document}


\title{Exact solution of the van der Waals model in the critical region}

\author{Adriano Barra$^{\footnotesize{\;1)}}$  and   Antonio Moro$^{\footnotesize{\;2)}}$ \\
~\\
\footnotesize{$^{1)}$Dipartimento di Fisica, Sapienza Universit\`{a} di Roma, Rome, Italy}\\
\footnotesize{$^{2)}$Department of Mathematics and Information Sciences, University of Northumbria at Newcastle upon Tyne, UK }}

\maketitle

\date{~}

\begin{abstract}
The celebrated van der Waals model describes simple fluids in the thermodynamic limit and predicts the existence of a critical point associated to the gas-liquid phase transition. However the behaviour of critical isotherms according to the equation of state, where a gas-liquid phase transition occurs, significantly departs from experimental observations. The correct critical isotherms are heuristically re-established via the Maxwell equal areas rule. A long standing open problem in mean field theory is concerned with the analytic description of van der Waals isotherms for a finite size system that is consistent, in the thermodynamic limit, with the Maxwell prescription.
Inspired by the theory of nonlinear conservation laws, we propose a novel mean field approach, based on statistical mechanics, that allows to calculate the van der Waals partition function for a system of large but finite number of particles $N$. Our partition function naturally extends to the whole space of thermodynamic variables, reproduces, in the thermodynamic limit $N\to \infty$, the classical results outside the critical region and automatically encodes Maxwell's prescription. 
We show that isothermal curves evolve in the space of thermodynamic variables like nonlinear breaking waves and the criticality is explained as the mechanism of formation of a classical hydrodynamic shock.  \\


\end{abstract}

\maketitle

\section{Introduction}
\begin{figure}[h]
\begin{center}
\includegraphics[scale=.35]{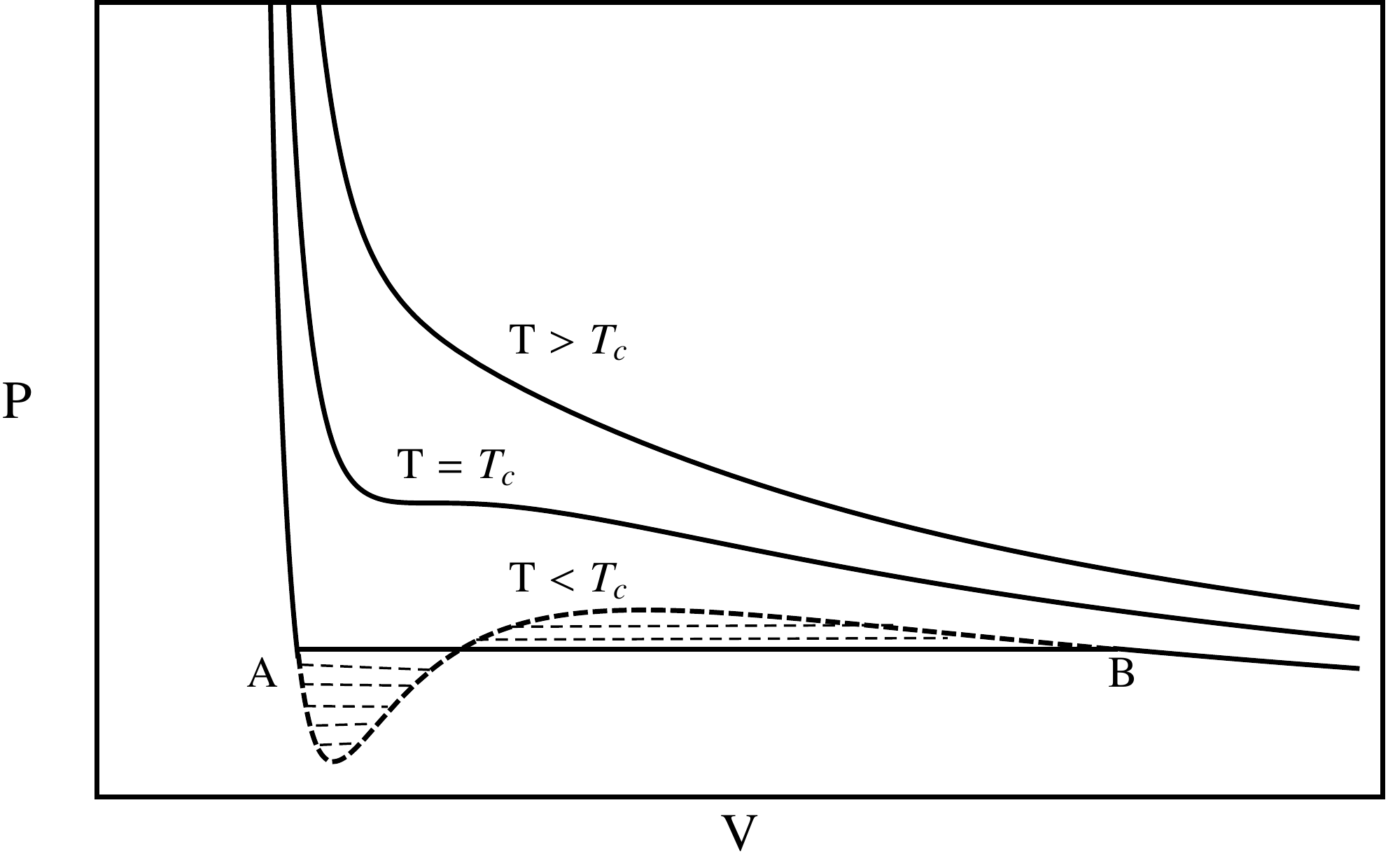}
\caption{Real gas isothermal curves: within the critical region, between the points $A$ and $B$ the behavior predicted by the van der Waals equation (dashed line) departs from the experimentally observed (solid line). The actual critical isotherms are constructed starting from the theoretical ones via the Maxwell equal areas rule.}
\label{fig:G1a}
\end{center}
\end{figure}

In late 19th century, due to the outstanding contribution of its founding fathers,  Boltzmann, Maxwell and Gibbs, Statistical  Thermodynamics has been successfully introduced as the general conceptual framework for understanding equilibrium thermodynamic phenomena by means of statistical mechanics~\cite{yanes,kittel}.
\newline
The early success of the kinetic theory of gases and the discovery of the mean field approach for the derivation of the classical van der Waals equation~\cite{stanley} nurtured the hope that an equally neat and clear description of critical phenomena would be as effective as it was away from the critical region. Although second order phase transitions, as for example the ergodicity breakdown of real gases at the triple point,  have nowadays been completely framed into a rigorous scaffold~\cite{gallavotti} - also partially guided by techniques from the near field theory~\cite{parisi,kardar} - first order phase transitions, such as gas-liquid phase transitions, turned out to be more elusive.
\newline
Indeed, in spite of the accuracy of the celebrated van der Waals equation for the description of real gases, the behavior predicted within the critical region, where a real gas turns into a liquid, significantly departs from the experimental observations. Fig.\ref{fig:G1a} shows the typical isothermal curves of a real gas: above the critical temperature (for $T>T_{c}$) the pressure decreases as a strictly monotonic function of the volume; the critical point corresponds to the temperature $T =T_{c}$ where the critical isotherm develops an inflection point; below the critical temperature (at $T<T_{c}$) real isotherms are constant within a certain volume interval in spite of the oscillating behavior predicted by the classical van der Waals equation.
Interestingly, the behavior of real isothermal curves within the critical region turns out to be intimately connected to the theoretical one via the celebrated Maxwell rule stating that the constant pressure plateau is placed in such a way it cuts lobes of equal areas on the associated van der Waals isotherm. As it is well known, the Maxwell rule corresponds to the condition of thermodynamic equilibrium such that, below the critical temperature, the Gibbs free energy develops two minima of equal value~\cite{Callen}. \\
The remarkable validity, although heuristic, of Maxwell's approach stimulated countless studies aimed at a rigorous statistical mechanical description of first order phase transitions as for instance in the works of Lebowitz and Penrose~\cite{lebowitz} and van Kampen~\cite{Kampen}, where large classes of pairwise interaction potentials for particles (continuous and hard-sphere-like respectively) are considered or the work by Griffiths~\cite{griffiths} that focusses on the study of analyticity properties of thermodynamic functions.
\newline
Alternative methods to analyze phase transitions have also been developed based on macroscopic approaches to thermodynamics. For instance, the Landau theory allows to construct suitable asymptotic expansions of the free energy in the order parameters to obtain information of the critical exponents in the vicinity of the critical point (see e.g.~~\cite{Toledano}); the Widom  approach relies on the construction of effective free energy functions based on the analysis of their scaling properties~\cite{Widom}. Further recent developments in this direction led to the formulation of the thermodynamic limit as the semiclassical limit of nonlinear conservation laws where phase transitions are associated to shock solutions of a hyperbolic nonlinear PDE in the class of conservation laws~\cite{sumrule,Genovese,moro2,noi}. Such nonlinear PDEs can be also derived in mean field theories from the analysis of differential identities of the free energy as showed in~\cite{sumrule,Genovese,newman2,brankov1, brankov2} for the Curie-Weiss and the Sherrington-Kirkpatrick models, or from the analysis of thermodynamic Maxwell relations as showed in~\cite{moro1,moro2} for the van der Waals model.
Both the microscopic statistical mechanical approach, via the study of correlation functions asymptotics, and the macroscopic thermodynamic approach, based on the expansion of the free energy in the vicinity of the critical point, show the intimate connection with the singularity and catastrophe theory - since the very first pioneering contributions by Arnold - and the Hopf bifurcation theory (see e.g.~\cite{Arnold}).

Despite the numerous progresses made in understanding phase transitions in a variety of contexts, from thermodynamics to classical and quantum field theory \cite{zamponi1,zamponi2},  or complex and biological systems \cite{mezard,MPV}, and the discovery of their intrinsic universality, a global analytical description of phase transitions for the van der Waals gas is still missing. In this work, inspired by the theory of nonlinear PDEs, in the class of nonlinear conservation laws, we propose a novel method such that given an equation of state assumed to be accurate outside the critical region allows to construct a partition function for a finite number of particles $N$ that is valid in the whole space of thermodynamic variables including the critical region. This partition function  automatically encodes the Maxwell equal areas rule.
\newline
Based on the mean field assumption that configurations of equal volume are equally weighted, we obtain the general functional form of the partition function. For finite $N$, the partition function is, as expected, analytic in the space of thermodynamic variables but it develops a singularity in the thermodynamic limit $N\to \infty$. We use the Laplace method for the asymptotic evaluation of the partition function for large $N$ with the constraint  that above the critical point, where the Laplace integral admits one single critical point, the leading asymptotics of the volume expectation value satisfies the classical van der Waals equation. Remarkably, this condition allows to fix uniquely the functional form of the partition function in such a way that the logarithm of the probability density gives the correct Gibbs free energy of the van der Waals model above the critical point. Finally, we prove that in the critical region, defined as the region in the space of thermodynamic variables where the Laplace integral admits multiple critical points, the leading asymptotics for the volume develops a discontinuous behavior, providing the exact analytical description of the first order phase transition.


\section{The model.} 
Let us consider a fluid of $N$ identical particles of mass $m$, whose centre of mass is fixed at the origin of the reference frame, described by the Hamiltonian of the form
\begin{equation}
\label{Hamilt_gen}
H_{N} = \sum_{l=1}^{N} \frac{{\bf p}_{l}^{2}}{2 m} - \frac{1}{2} \sum_{l,m =1}^{N} \psi ({\bf r}_{l}, {\bf r}_{m}) + P  v({\bf r}_{1},\dots, {\bf r}_{N}),
\end{equation}
where ${\bf p}_{l}$ is the momentum of the particles $\psi({\bf r}_{l}, {\bf r}_{m})$ is a two-body interaction potential, and the last term models the interaction with an external field where $P>0$ is a real positive  mean field coupling constant and the volume $v({\bf r}_{1},\dots, {\bf r}_{N})$ is defined as the minimum convex hull associated to the configuration $\{{\bf r}_{1},\dots, {\bf r}_{N}\}$. The partition function is given by the standard formula for a canonical ensemble 
\[
{\cal Z} = \int d^{N}{\bf p}_{i} d^{N}{\bf r}_{i} e^{-\beta H_{N}}
\]
where $\beta = (K_{B} T )^{-1}$ with $K_{B}$ the Boltzmann constant and $T$ the temperature. Let us observe that fixing the centre of mass breaks the translational invariance of the Hamiltonian~(\ref{Hamilt_gen}) that otherwise would lead to a divergent partition function ${\cal Z}$. Integration over the moment variables ${\bf p}_{l}$ returns ${\cal Z} = (2 \pi m K_{B} T)^{3N/2} Z$ where
\begin{equation}
\label{partition}
Z = \int  d^{N} {\bf r}_{i} \; \exp \left [ N \left( \frac{t}{2}  \sum_{l,m} \psi \left({\bf r}_{l}, {\bf r}_{m}  \right ) + x v \right) \right].
\end{equation}
The rescaled variables $t = 1/(N K_{B} T)$ and $x = - P/(N K_{B} T)$ introduced above will allow us to define the thermodynamic limit and the choice of the notation emphasizes the formal analogy between the Gibbs free energy and the Hamilton-Jacobi function of the associated mechanical problem (see e.g.\cite{Genovese}).
Let us introduce the density of free energy 
$$\alpha_{N} = -N^{-1} \log {\cal Z}.$$ 
This quantity is sometimes also referred to as {\em mathematical pressure}, see e.g. \cite{GRS}. The expectation value of a given observable ${\cal O}$ is defined in the usual manner, i.e. $$\av{{\cal O} } = {\cal Z}^{-1}\int  d^{N}{\bf r} \; {\cal O}  \; e^{-\beta H_{N}}.$$ 
In particular, let us observe that $\av{v}  = - \partial \alpha_{N}/\partial x $.



The thermodynamic regime is defined as the large particles limit $N \sim N_{A}$ where $N_{A} \simeq  6.022 \times10^{23}$ is Avogadro's number. In particular, for $n$ moles of  a gas of molecules of hard core volume $b_{0}$ we have
\[
N =n N_{A} \qquad  N K_{B} = n R \qquad N b_{0} = n b_{m}
\]
 where $R = N_{A} K_{B} \simeq 8.31 J/mol K$ is the gas constant and $b_{m} = N_{A} b_{0}$ is the molar hard core volume. Hence, the gas constant $R$ defines the typical scale for the variables $x$ and $t$.

We now assume that for fixed values of the variables $x$ and $t$,  configurations of equal volume occur with the same probability density, so that there exists a probability measure $\mu(v)$ such that the partition function~(\ref{partition}) is of the form 
\begin{equation}
\label{Zmeasure}
Z =\int_{b}^{\infty} d\mu(v)
\end{equation}
where $b = n b_{m}$ is the total hard core volume. This assumption gives a nonlinear generalization of the standard mean field approximation introduced for the statistical mechanical derivation of the van der Waals equation of state (see e.g.\cite{stanley}). We also note that, from a formal perspective, this ansatz is equivalent to the request that the moments $\av{v^{n}}$ for the model~(\ref{Hamilt_gen}) are such that the measure $\mu(v)$ is the solution to the {\it Stieltjes moments problem}~\cite{Akhiezer}, that is 
\[
\av{v^{n}} =   \int_{b}^{+ \infty} v^{n} d \mu(v). 
\]
Expressing the differential as $d\mu(v) = \mu'(v) dv $, the function $\mu'(v)$ gives the weight associated to a given volume configuration that, for fixed values of $x$ and $t$, is the same for all configurations of equal volume. As for the canonical ensemble the logarithm of the probability density in~(\ref{partition}) is linear in the variables $x$ and $t$ (this ensures entropy maximization at equilibrium), we have that the probability density $\mu'(v)$ is such that $\log \mu'(v) = N \left(x v +\frac{1}{2} t \phi(v)  + \sigma(v) \right)$
for certain functions $\phi(v)$ and $\sigma(v)$. Hence, the partition function takes the form
\begin{equation}
\label{part_mom}
Z = \int_{b}^{+\infty} d \mu(v) = \int_{b}^{+\infty} e^{N (x v +\frac{t}{2} \phi(v)  + \sigma(v))} \; dv.
\end{equation}
In the following  we will prove that the functions $\phi(v)$ and $\sigma(v)$ can be uniquely determined by the request that the expectation value $\av{v}$ evaluated, away from the critical region, according to the partition function~(\ref{part_mom}) satisfies, in the thermodynamic limit, the celebrated van der Waals equation
\begin{equation}\label{classical_vdW}
\left (P + \frac{a}{v^2} \right) \left (v - b \right) = n R T.
\end{equation}
We should stress that, as discussed in details below, the assumption about the existence of the measure in~(\ref{Zmeasure}) is a strong enough information to fix uniquely the functional form of  $\varphi(v)$ and $\sigma(v)$ with no further specifications on the two-body potential $\psi({\bf r}_{l},{\bf r}_{m})$. However, we find instructive to present an heuristic phenomenological construction for a class of two-body nearest neighbourhood potential depending on the distance of the form $\psi({\bf r}_{l},{\bf r}_{m}) = \psi(r_{lm})$, where $r_{lm} = |{\bf r}_{l} - {\bf r}_{m} |$. In the thermodynamic limit, we can assume that for a given equilibrium configuration particles are on average approximately equidistant,  i.e. $r_{lm} \simeq \bar{r}$ where $\bar{r}$ is the mean distance, then
\[
\sum_{l,m} \psi(r_{lm}) \simeq \sum_{l,m} \psi(\bar{r}) \sim  N \psi(\bar{r}).
\]
We used the fact that the number of nearest neighbourhood pairs is of order $N$. Let us consider for example an effective electric potential energy
\[
\psi(\bar{r}) \simeq \frac{1}{N^{\beta}} \frac{q^{2}}{4 \pi \epsilon_{0} \bar{r}^{\alpha}} \simeq \frac{q^{2}}{4 \pi \epsilon_{0} v^{\alpha/3}}
\]
where we observed that the mean distance for a given volume configuration is related to the volume per particle by the relation $\bar{r} \simeq (v/N)^{1/3}$ and the exponent $\beta =\alpha/3$ is chosen to ensure the linear extensivity of the potential term in~(\ref{Hamilt_gen}). More in general we assume that in the thermodynamic limit we can write
\[
\sum_{l,m} \psi({\bf r}_{l},{\bf r}_{m}) \simeq \phi(v),
\]
that is the potential energy can be expressed as a function of the volume.
Under this assumption, we observe that the partition function~(\ref{partition}) can be equivalently written as
\[
Z = \int_{b}^{\infty} dv \; {\cal D}_{N}(v) e^{N \left (\frac{t}{2} \phi(v) + x v \right)}
\]
where 
\[
{\cal D}_{N}(v) = \int d^{N}{\bf r}_{i} \delta \left (v - v({\bf r}_{1},\dots,{\bf r}_{N}) \right)
\]
gives the number of configurations of the prescribed volume $v$.
A direct comparison with the formula~(\ref{part_mom}) leads us to the natural interpretation, for large $N$, of the function $\sigma(v)$ as the {\it configurational entropy} of the system i.e.
\[
\sigma(v) \simeq \frac{1}{N} \log {\cal D}_{N}(v).
\]
\newline
Let us now proceed by evaluating the leading order asymptotics, for $N\to \infty$, of the partition function~(\ref{part_mom}) in the region of thermodynamic variables $x$ and $t$ where $\log \mu'(v)$ admits one single critical point. Laplace's formula gives
\begin{equation}
\label{part_laplace}
Z \simeq \sqrt{\frac{2 \pi}{N \alpha''(v^{\star})}} \; e^{- N \alpha(v^{\star})}, \qquad N \to \infty
\end{equation}
where $\alpha(v) =  - x v -  t \phi(v)/2  -   \sigma(v)$ and $v^{\star}(x,t)$ is a stationary point for the potential $\alpha(v)$ such that $\alpha'(v^{\star}) = 0$, i.e.
\begin{equation}
\label{G_stat}
x + \frac{t}{2} \phi'(v^{\star}) + \sigma'(v^{\star}) = 0.
\end{equation}
In particular, formula~(\ref{part_laplace}) implies that $\alpha = \lim_{N\to \infty} \alpha_{N}$.
Identifying the external field constant $P$ in the Hamiltonian~(\ref{Hamilt_gen}) with the physical pressure in Eq.~(\ref{classical_vdW}) and choosing 
\begin{subequations}
\label{matching}
\begin{align}
\label{matching1}
\phi(v) =&2 a /\av{v}\\ 
\label{matching2}
\sigma(v) =&  \log \left(\av{v} - b \right)
\end{align}
\end{subequations}
Eq.~(\ref{G_stat}) coincides with the van der Waals equation~(\ref{classical_vdW}). 
We also note that the asymptotic matching condition~(\ref{matching2}) allows us to evaluate the  function ${\cal D}_{N}(v)$ for large $N$ that as expected is
\[
{\cal D}_{N}(v) \simeq \left (\av{v} - b \right )^{N}
\] 
The prescriptions~(\ref{matching}) are, according our procedure, the necessary matching conditions that uniquely fix the partition function~(\ref{part_mom}) consistently  with the van der Waals equation of state, which is assumed to be accurate above the critical region. We note that $\alpha(v) = G/T$, where $G$ is the Gibbs free energy density of the van der Waals model. A direct calculation shows that the partition function so obtained
\begin{equation}
\label{partKG}
Z = \int_{b}^{\infty} e^{N \left(x v + t \frac{a}{v} + \log \left(v - b \right) \right)} \; dv
\end{equation}
satisfies the Klein-Gordon equation in the light-cone variables
\begin{equation}
\label{KG}
\frac{\partial^{2} Z}{\partial x \partial t} =  N^{2} a Z.
\end{equation}
Let us also observe that the integral expression~(\ref{partKG}) can be explicitly evaluated at finite $N$ for $t =0$ and gives $\av{v}(x,0) =  b - 1/x - 1/N x$ that coincides with the equation of state~(\ref{classical_vdW}) in the limit $T \to \infty$.

\begin{figure}
\begin{center}
\includegraphics[scale=.4]{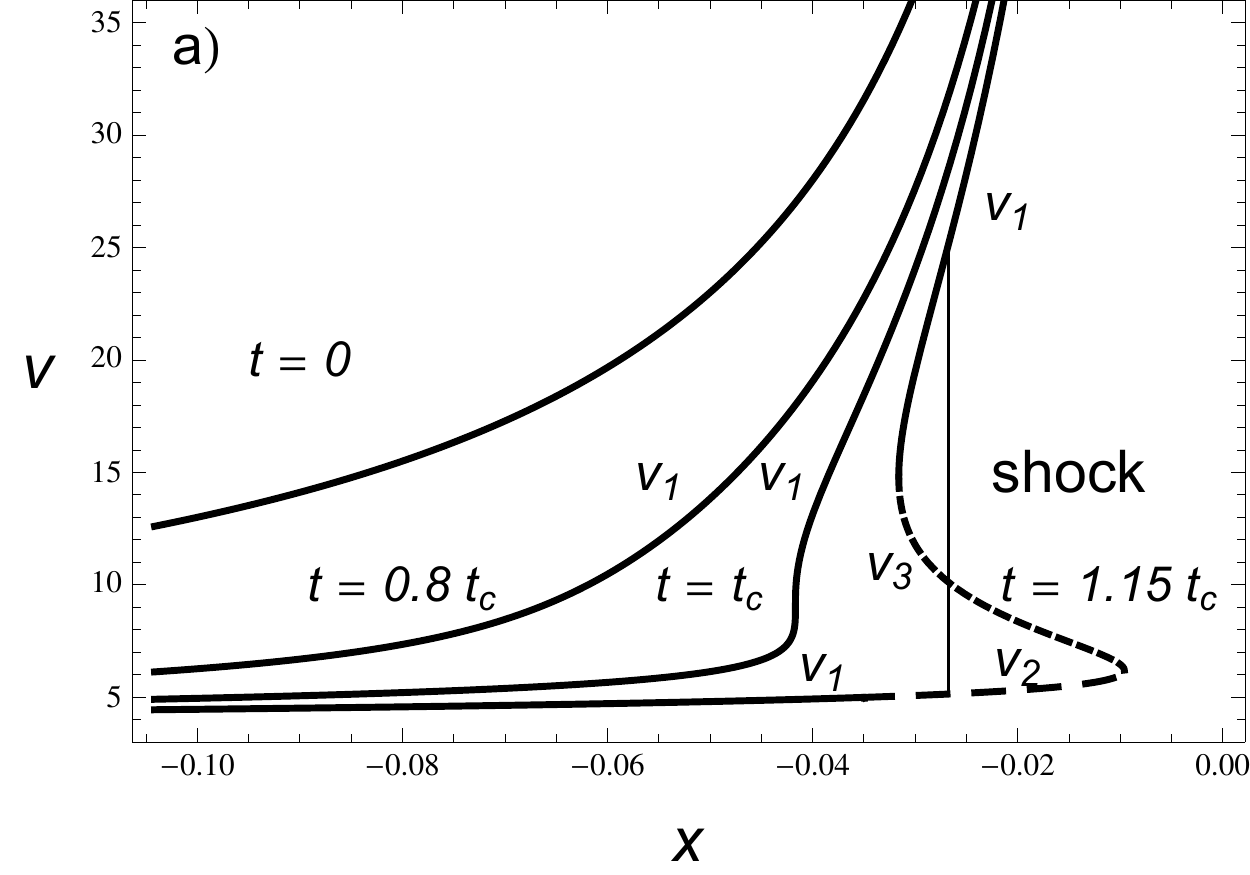} \includegraphics[scale=.4]{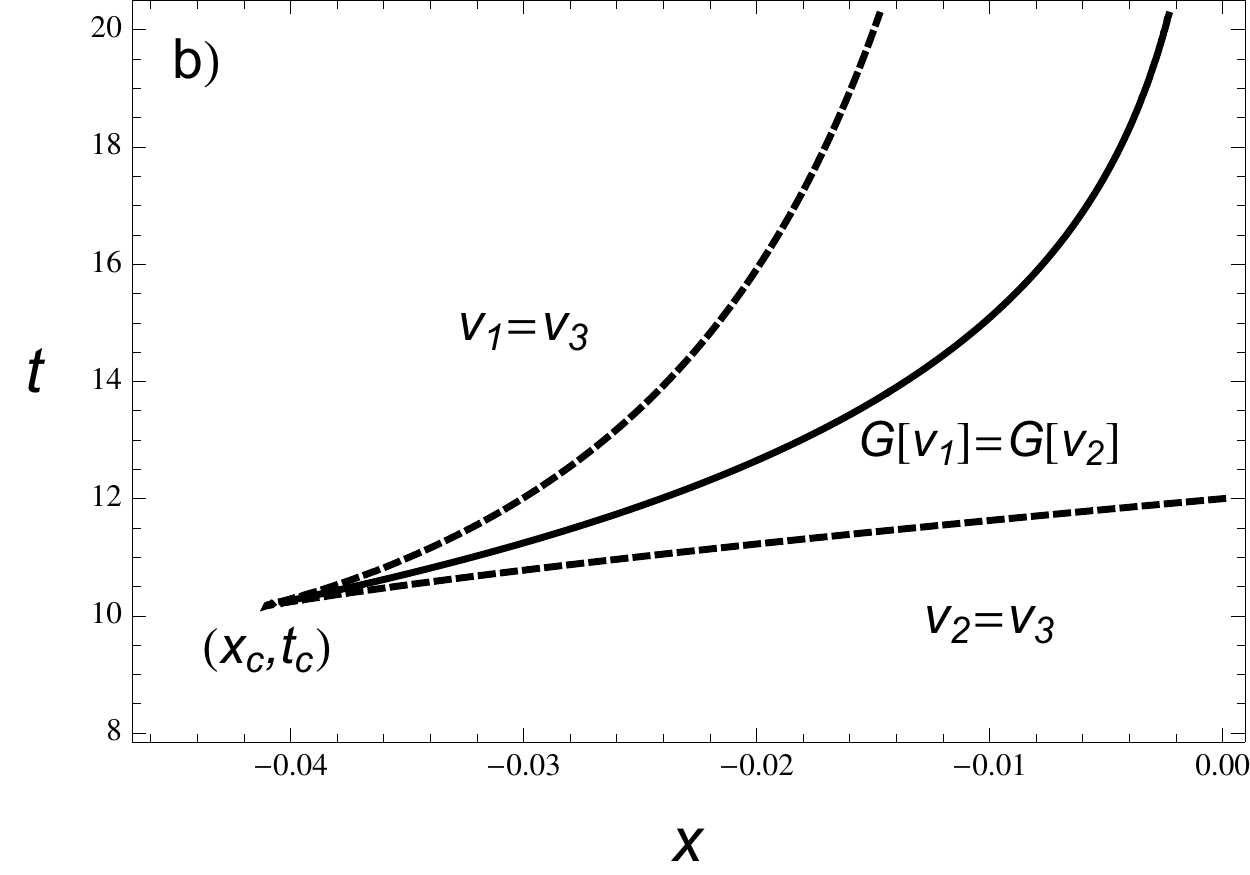}
\caption{a) van der Waals isothermal curves (for the choice of parameters $a=1$ and $b=3$) above and below the critical temperature $T_{c} = (N K_{B} t_{c})^{-1}$ .
b) Shock trajectory (solid line) and critical sector (delimited by the dashes lines) associated to multivalued isotherms.
}
\label{fig:G1}
\end{center}
\end{figure}

Using the self-consistency equation
\[
\av{v} = N^{-1} Z^{-1} \partial Z /\partial x
\] 
the Klein-Gordon equation~(\ref{KG}) implies that the volume density satisfies the nonlinear viscous conservation law
\begin{equation}
\label{vfulleq}
\der{\av{v}}{t} = \der{}{x} \left(\frac{a}{\av{v}} + \frac{1}{N} \der{\log \av{v}}{t} \right)
\end{equation}
of the type studied in~\cite{ALM} and that is related to the viscous analog of the Camassa-Holm equation. In the thermodynamic limit, above the critical temperature where the gradient of $\av{v}$ is bounded, the term of order $O(N^{-1})$ in~(\ref{vfulleq}) is negligible and the volume density satisfies the Riemann-Hopf type equation
\[
\partial_{t}\av{v} = \partial_{x} (a/\av{v})
\] 
whose solution develops a gradient catastrophe in finite ``time'' $t$. As illustrated in Fig.~\ref{fig:G1}.a, the volume $\av{v}$ evolves in the space of thermodynamic parameters just like a nonlinear hyperbolic wave and the gradient catastrophe is associated to the critical point $x_{c} = - 1/8b$, $t_{c} = 27b/8a$, $v_{c} = 3 b$. Beyond the critical time $t_c$, the physical solution develops a shock discontinuity, corresponding to a first order phase transition, whose position at fixed $t > t_{c}$ is determined by the equal area rule and its speed $U$ is given the Rankine-Hugoniot condition 
$$U = -(a/v_l-a/v_r)/(v_l-v_r) = a/(v_l v_r)>0,$$
where $v_{l}$ and $v_{r}$ are the limiting values of $\av{v}$ respectively to the left and to the right of the jump. It was observed in~\cite{moro1} that the Rankine-Hugonoit condition is equivalent to the Clausius-Clapeyron equation implying that the shock speed is proportional to the latent heat associated to the first order phase transition and the trajectory of the shock is interpreted as the coexistence curve of the gas-liquid phase as shown in Fig.{\ref{fig:G1}.b}. Such connection between phase transitions and scalar shock waves was first observed in the context of magnetic models (see e.g.~\cite{noi} and reference therein), and in the classical thermodynamic setting in~\cite{moro2,moro1}  where the notion of universality has been also discussed.

It is interesting to compare the mean field partition function~(\ref{partKG}) associated to the model Hamiltonian~(\ref{Hamilt_gen}) and the equation of state obtained according to the standard canonical ensemble formalism. For the sake of simplicity let us consider an ideal gas of non-interacting particles of hard core volume $b$. 
In this case the $\phi(v) = 0$ and the coupling constant $P$ models the interaction with the external environment. We should stress that in the present formalism the gas does not occupy a prescribed volume, but the mean field partition function~(\ref{partKG}) accounts for all possible gas configurations over the whole space. Evaluating explicitly the formula~(\ref{partKG}) we obtain
\[
Z = \frac{(N-1)!}{N^N} \left(-\frac{1}{x} \right)^{N+1} e^{N b x}.
\]
The equation of state for the gas of $N$ particles is given, in full analogy with the case of mean field spin systems (see e.g.~\cite{Genovese}) via the self-consistency equation
\[
\av{v} = \frac{1}{N} \frac{1}{Z} \der{Z}{x} = b - \frac{N+1}{N x}
\]
or equivalently
\[
(\av{v} - b) P = (N+1) K_{B} T.
\]
In the thermodynamic regime $N \simeq n N_{A}$ we obtain the well known ideal gas equation of state
\begin{equation}
\label{idealeq}
(\av{v} - b) P = n R T,
\end{equation}
which allows us to identify the coupling constant $P$ with the physical pressure that plays the role of the external magnetic field in spin systems. \\  Within the canonical formalism the ideal gas constituted by a fixed number of particles $N$ of Hamiltonian
\[
H = \sum_{l=1}^{N} \frac{{\bf p}_{l}^{2}}{2 m}
\]
 is assumed to be in equilibrium with an external reservoir and occupy a prescribed volume, say $V$. The partition function  is given by
\[
{\cal Z} = \int d^{N}{{\bf p}_{i}} \; d^{N}{{\bf r}_{i}} \; e^{-H/K_{B} T} = (2 \pi m K_{B} T)^{3N/2} \left(\int d {\bf r} \right)^{N} = (2 \pi m K_{B} T)^{3N/2} {(V-b)}^{N}.
\]
The pressure
\[
P = - \der{F}{V}
\]
is then defined in terms of the Helmholtz free energy $F(V,T) = - K_{B} T  \log {\cal Z}$ gives the equation of state in the form~(\ref{idealeq}). Unlike the canonical formalism the mean field Hamiltonian~(\ref{Hamilt_gen}) encodes the boundary condition weighing different volume configurations that are intrinsically defined via the minimum convex hull. As a consequence the number $N$ associated to the spatial scale of the system takes into account of finite size effects, which are important near the criticality, in a consistent manner with the Maxwell rule.

\vspace{.2cm}
{\bf Remark.} We observe that the procedure presently described, that allows us to extend the van der Waals equation of state to the critical region, can straightforwardly be generalised to the class of equations of state obtained from (large volumes) virial expansions of the form (see e.g.~\cite{Landau})
\[
P = \frac{n R T}{v} \left( 1 + \frac{B_{1}(T)}{v} + \frac{B_{2}(T)}{v^{2}} \dots \right)
\]
with 
\[
B_{i}(T) = \frac{\alpha_{i+1}}{2 n R T} + \beta_{i+1} \qquad i = 1,2,3,\dots
\]
where $\alpha_{i}$ and $\beta_{i}$ are real constants given by the large volume asymptotic expansion of the functions $\sigma(v)$ and $\phi(v)$ of the form
\begin{align*}
\phi(v) = - \sum_{k=1}^{\infty} \frac{\alpha_{k+1}}{k v^{k}} \qquad
\sigma(v)  = \log  V - \sum_{k=1}^{\infty} \frac{\beta_{k+1}}{k v^{k}}.
\end{align*}

\section{The critical region.} 
The subset of the space of thermodynamic variables $x$ and $t$ where the free energy $\alpha(v)$ admits multiple critical (stationary) points defines the critical region associated to the gas-liquid phase transition. In this case, the leading asymptotics at large $N$ of the  partition function~(\ref{partKG}) is given by the formula
\begin{equation}
\label{part_laplace_mult}
Z \simeq \sum_{i} \sqrt{\frac{2 \pi}{N \alpha''(v_{i})}} \; e^{- N \alpha(v_{i})}, \qquad N \to \infty
\end{equation}
where the sum runs over the local minima $v_{i} (x,t)$ of the free energy $\alpha(v)$.
Hence, consistently with the classical description of the van der Waals phase transition, below the critical temperature the Gibbs free energy develops three stationary points, two of which are local minima. In the limit $N \to \infty$ the leading contribution to the partition function is given by the point of local minimum $v_{m}$ such that $\alpha(v_{m}) \leq \alpha(v_{i})$, for all $i \neq m$. Hence, within the critical region, the solution is given by 
\[
\av{v} =  \lim_{N\to \infty}N^{-1} \partial_{x} \log Z = v_{m}
\]
where $v_{m}(x,t)$ is a root of the equation of state $\alpha'(v_{m}) = 0$ such that the Gibbs free energy has the lowest local minimum.
The subset of the $(x,t)$-plane, such that $\alpha$ takes two equal minima $\alpha(v_{i}(x,t)) = \alpha(v_{j}(x,t))$, represents the curve of resonance of the exponential contributions in~(\ref{part_laplace}) and identifies the shock line shown in Fig.~\ref{fig:G1}.b. As already known from the theory of classical shocks for the viscous Burgers equation, such resonance condition is equivalent to the equal areas rule~\cite{Whitham}.

\begin{figure}
\begin{center}
\vspace{.3cm}
\includegraphics[scale=.4]{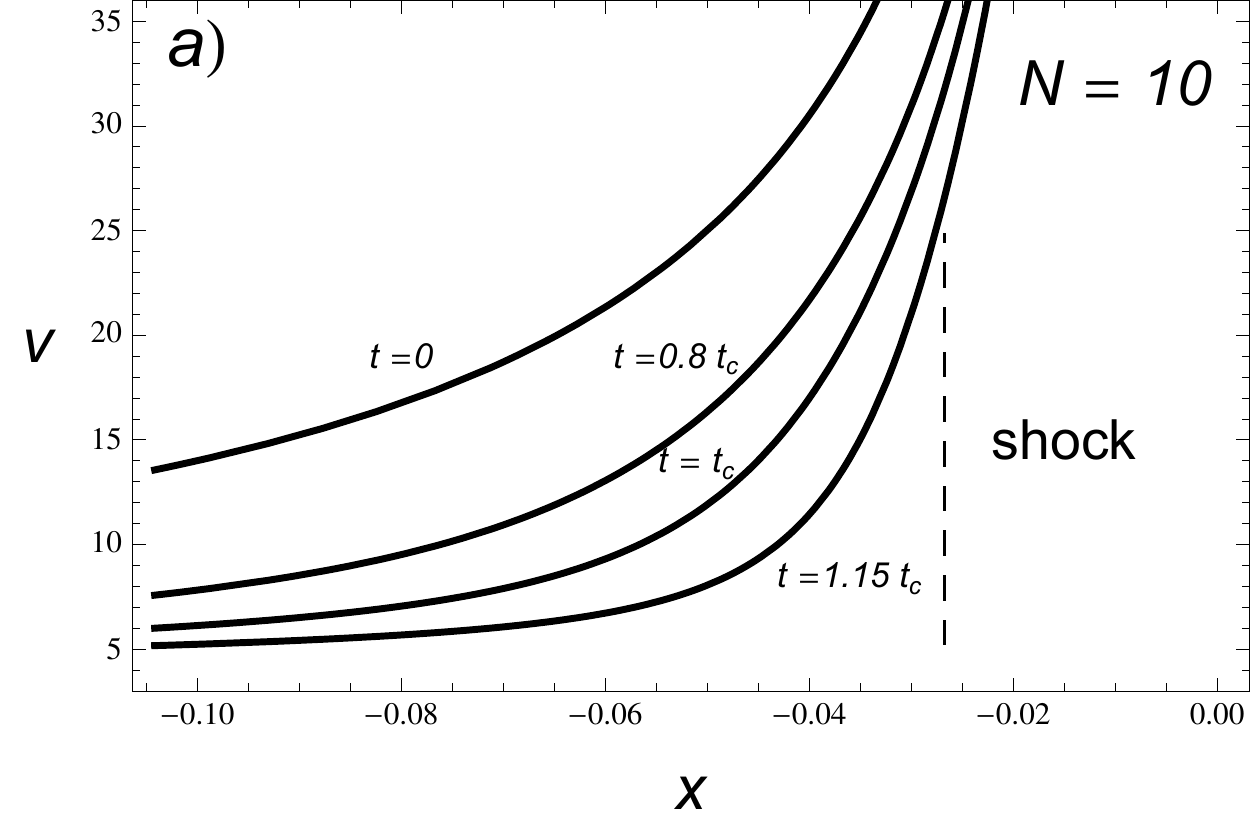} \includegraphics[scale=.4]{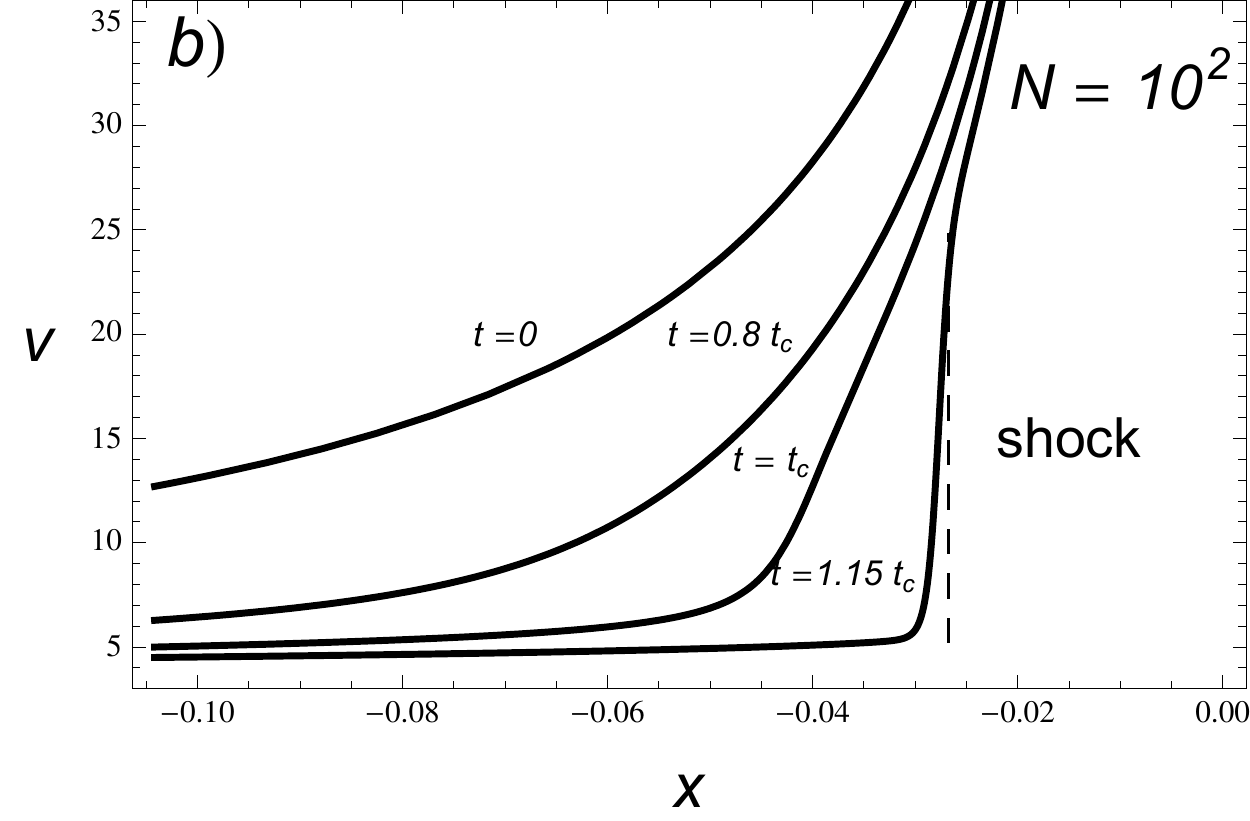} \\\includegraphics[scale=.4]{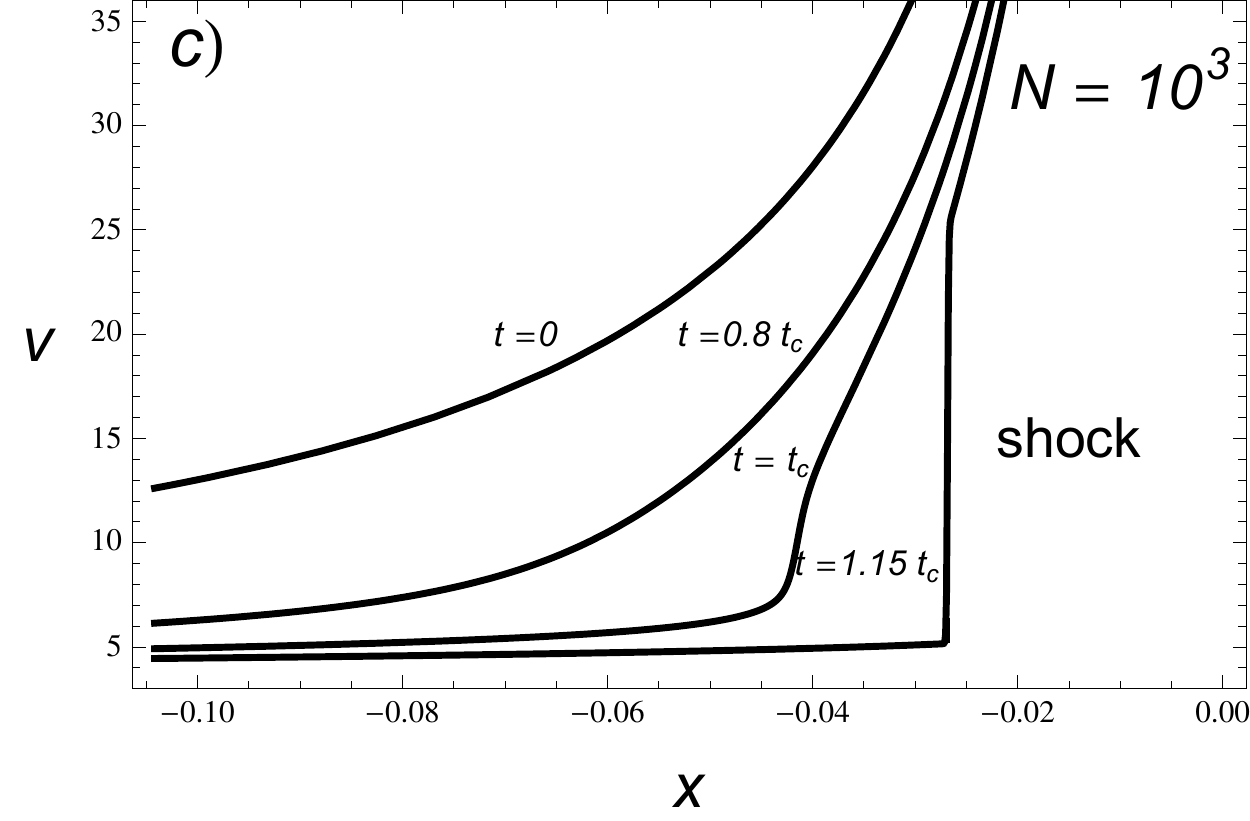} \includegraphics[scale=.4]{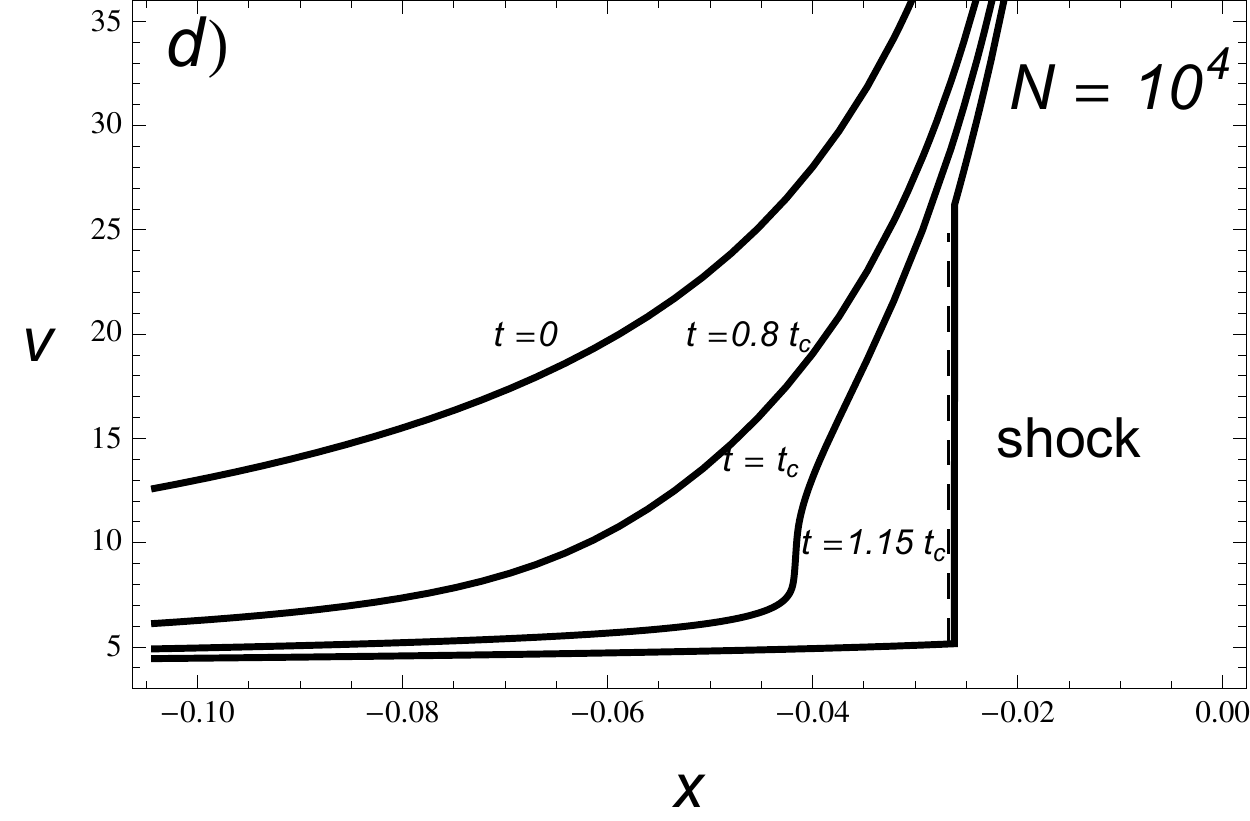}
\caption{Solution for different values of $N$ and comparison with the classical shock at $t = 1.15 t_{c} $ for the choice of parameters $a=1$ and $b=3$.}
\label{fig:G2}
\end{center}
\end{figure}
In Fig.\ref{fig:G2} we plot the isothermal curves evaluated using the partition function~(\ref{partKG}). As $N$ increases the exact isothermal curves develop an inflection point and rapidly converge to the asymptotic behavior predicted by the Laplace formula. We should emphasize that the partition function~(\ref{partKG}) provides a global description of isothermal curves in the space of thermodynamic variables and the description of the phase transition is apparently accurate already for relatively small $N \simeq 10^{4}$ if compared with Avogadro's number.
The formula~(\ref{partKG}), provides an explicit description of how finite size effects play the role of a singularity resolution mechanism. It also gives a statistical mechanical based interpretation the results obtained in~\cite{moro1}
that allow to identify the multi-scale regime characterizing the universal local form of the equation of state
\[
v = v_{c} + N^{-1/4} u \left( \frac{x-x_{c} + a (t-t_{c})/v_{c}^{2}}{N^{-3/4}}, \frac{t-t_{c}}{N^{-1/2}} \right)
\]
where 
\[
u(\xi,\tau)  = -2 \der{\log \Lambda}{\xi} (\xi,\tau),
\]
$\Lambda(\xi,\tau)$ is the Pearcey integral
\[
\Lambda(\xi,\tau) = \int_{-\infty}^{\infty}   e^{-\frac{1}{8} (z^{4} - 2 \tau z^{2} + 4 \xi z)} \; dz
\]
and $(x_{c},t_{c},v_{c})$ are the coordinates of the critical point as evaluated above.

\section{Concluding remarks.} 
This work shows how our approach based on the combination of Statistical Mechanics and nonlinear PDEs theory provides us with a novel and powerful tool to tackle phase transitions.
This method leads to solution of perhaps the most known test-case that exhibits a first order phase transition (semi-heuristically described) such as the van der Waals model. In particular we have obtained the first global mean field partition function (eq.~(\ref{partKG})), for a system of finite number of particles. The partition function is a solution to the Klein-Gordon equation, reproduces the van der Waals isotherms away from the critical region and, in the thermodynamic limit $N\to \infty$ automatically encodes the Maxwell equal areas rule. The approach hereby presented is of remarkable simplicity, has been successfully applied to spin~\cite{newman2,brankov1,brankov2,Genovese, noi} and macroscopic thermodynamic systems~\cite{moro1, moro2}  and can be further extended to include the larger class of models admitting partition functions of the form~(\ref{part_mom}) to be used to extend to the critical region general equations of state of the form~(\ref{G_stat}) including a class virial expansions.\\

\noindent {\bf Acknowledgements.} AB has been partially supported by Progetto giovani GNFM-INdAM 2014 "Calcolo parallelo molecolare" and AM has been partially supported by Progetto giovani GNFM-INdAM 2014 "Aspetti geometrici e analitici dei sistemi integrabili".



\end{document}